\documentclass[conference]{IEEEtran}
\IEEEoverridecommandlockouts
\pdfoutput=1

\usepackage{cite}
\usepackage[cmex10]{amsmath}
\usepackage{amsmath,amssymb,amsfonts}
\usepackage{multicol}
\usepackage{graphicx}
\usepackage{caption}
\usepackage{algorithm}
\usepackage{algpseudocode}
\usepackage{textcomp}
\usepackage{tabularx}
\usepackage{comment}
\usepackage{xcolor}
\usepackage{array}

\begin{document}
%
\title{Mixed-Integer Optimization for Bio-Inspired Robust Power Network Design}



\author{Hao Huang, \emph{Student Member, IEEE,}\IEEEauthorrefmark{1} \IEEEauthorblockN{Varuneswara Panyam\IEEEauthorrefmark{2} , Mohammad Rasoul Narimani \emph{Member, IEEE,}\IEEEauthorrefmark{1}\\Astrid Layton \IEEEauthorrefmark{2}, Katherine R. Davis, \emph{Senior Member, IEEE,}\IEEEauthorrefmark{1}}
\IEEEauthorblockA{\IEEEauthorrefmark{1}Department of Electrical and Computer Engineering Texas A\&M University, College Station, TX, USA
\\ \IEEEauthorrefmark{2}Department of Mechanical Engineering Texas A\&M University, College Station, TX, USA
\\Email: hao\_huang@tamu.edu, vpanyam@tamu.edu, narimani@tamu.edu, alayton@tamu.edu, katedavis@tamu.edu}
}

\IEEEoverridecommandlockouts
\IEEEpubid{\makebox[\columnwidth]{978-1-7281-8192-9/21/\$31.00~\copyright2021 IEEE\hfill} \hspace{\columnsep}\makebox[\columnwidth]{ }}

\maketitle

\begin{abstract}

Power systems are susceptible to natural threats including hurricanes and floods. Modern power grids are also increasingly threatened by cyber attacks. Existing approaches that help improve power system security and resilience may not be sufficient; this is evidenced by the continued challenge to supply energy to all customers during severe events. This paper presents an approach to address this challenge through bio-inspired power system network design to improve system reliability and resilience against disturbances.
Inspired by naturally robust ecosystems, this paper considers the optimal ecological robustness that recognizes a unique balance between pathway efficiency and redundancy to ensure the survivability against disruptive events for given networks. This paper presents an approach that maximizes ecological robustness in transmission network design by formulating a mixed-integer nonlinear programming optimization problem with power system constraints. The results show the increase of the optimized power system's robustness and the improved reliability with less violations under $N-x$ contingencies.

 \end{abstract}



\begin{IEEEkeywords}
Power Networks Design, Mixed-Integer Nonlinear Programming, Power System Resilience
\end{IEEEkeywords}


\section{Introduction}

Power systems are large-scale geographically distributed critical infrastructures that support
other critical infrastructures and the functions of modern society, so 
resilience of these systems is a critical need. 
Their widespread presence 
inevitably
exposes them to natural disasters such as hurricanes, earthquakes, floods, etc. \cite{wang2015research} which can result in blackouts of different scales. 
The grid resilience report by the US National Academies calls for enhanced power system abilities to prepare for, endure, and recover from severe hazards~\cite{NationalAcademiesofSciencesEngineering2017}. The 2019 New York City Blackout, reportedly due to the malfunction of an aging transformer \cite{NYBlackoutCause}, motivates grid upgrades and construction. Modern power grids require not only advanced technology, but also reliable construction, to secure grid function and improve resilience. 
Power grid infrastructure
in developed countries have served the load for over a century. 
The challenge going forward
is thus how to invest, strengthen, and redesign power grid networks to improve resilience for increasing demand and dependence on electric energy. 

Power system resilience has been defined with time-dependent metrics capturing critical system degradation and recovery characteristics \cite{panteli2017metrics}. 
To enhance resilience, methods have been proposed for different event stages.
For example, Panteli \textit{et al} list several aspects of boosting power system resilience, such as enhanced situational awareness, deployment of renewables, network reconfiguration, and strengthening construction \cite{panteli2017power}. 
Risk to power system assets based on steady-state and transient analysis respectively 
along with cyber threats is prioritized in \cite{davis2016cyber,huang2018power} for improved cyber-physical situational awareness.
In \cite{khoshjahan2019harnessing}, Khoshjahan et al. formulate an optimization framework to enhance power grid flexibility with renewable energy. An algorithm for early detection and correction of power system insecurity at minimum cost has been proposed in \cite{gu2012early}. 
In any stage, 
a strategic network design lays the foundation for flexible operations and can itself mitigate some disturbances; 
we define this as \textit{robust network design}, and it is the focus of this paper.

Power network design necessitates considering realistic power system constraints \cite{rider2007power} 
and reliability guidelines, i.e., from 
the North American Electric Reliability Corporation (NERC). To deal with these considerations all together, different network expansion frameworks have been proposed. In \cite{kazerooni2010transmission}, Kazerooni \textit{et al} formulate the transmission network planning problem to minimize the sum of annual generator operating costs and annuitized transmission investment cost. To deal with the uncertainty of renewable generation and load, Jabr proposes a robust optimization approach for transmission network expansion planning \cite{jabr2013robust}. Moreover, the network structure of power grids draws significant attention from network and topological perspectives, which aim to analyze the robustness of power systems with complex network concepts and provide insights for robust power network design \cite{kocc2014topological,cuadra2015critical}. 
To provide a systematic resilient network design, it is important to have a valid benchmark for the problem formulation and analyses. Thus, in this paper, we refer to ecosystems, whose network structures have been shown to be sustainable over time while effectively withstanding disturbances, as our guideline for resilient network design, as described below.

\begin{figure}
\centering
\includegraphics[scale=0.3]{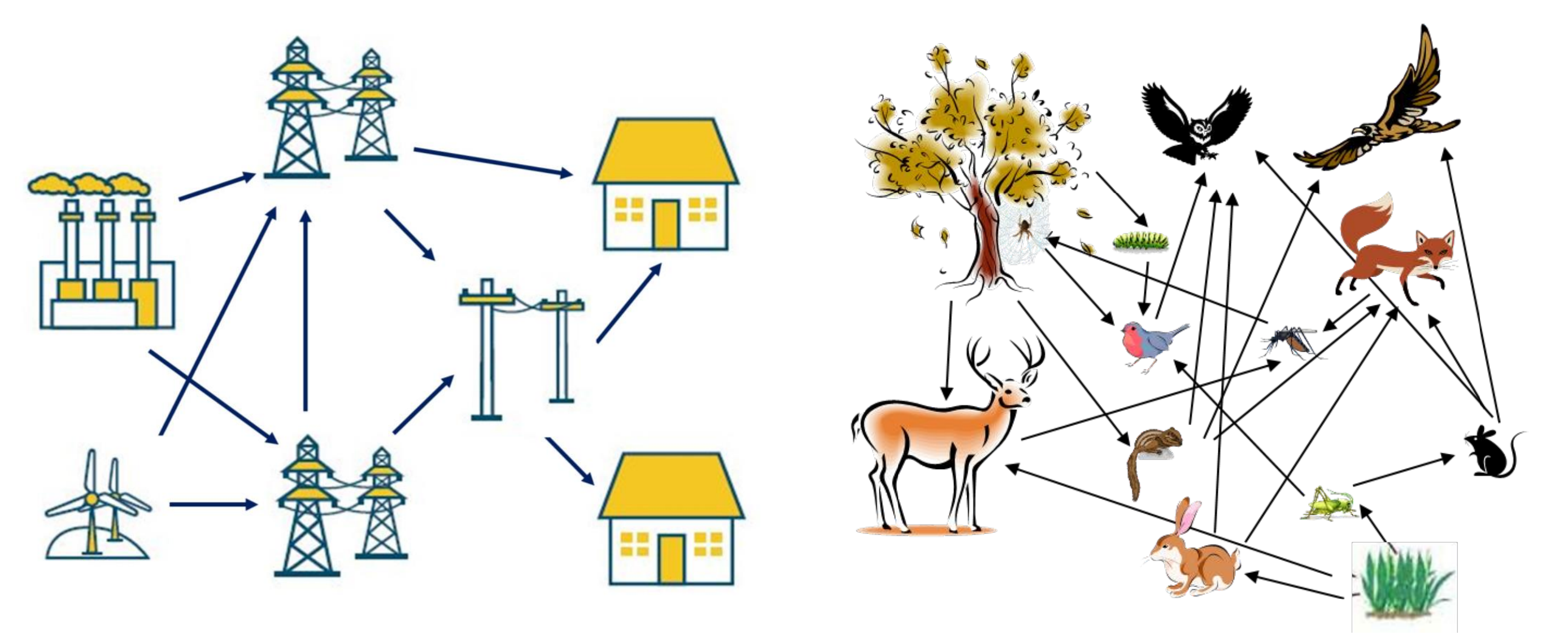}
\caption{A side-by-side illustration of the structurally similar topologies of a power grid (left) and a food web (right) \cite{panyam2019bioT}.}
\label{fig:web}
\end{figure}


As shown in Fig. \ref{fig:web}, the structure of food webs and power systems are similar. The actors and material exchange in food webs correspond to power system components and power transfer respectively. Biological ecosystems have developed over millions of years to survive disturbances, and they provide a benchmark for robust, sustainable, and efficient networks design. Ulanowicz \textit{et al} introduce ecological robustness to measure the potential of food webs to sustain over time \cite{ulanowicz2009quantifying}. 
In \cite{panyam2019bioT, panyam2019bioA}, Panyam \textit{et al.} introduce the bio-inspired robust power network design that applies ecological robustness to transmission network analyses and design based on network structure and real power flows. 
After applying ecological robustness as the objective to optimize the network structure, 
robustness is improved.
Compared with original networks, the bio-inspired power networks can withstand more high-impact events using an ?$N-x$? reliability evaluation criteria \cite{NERCTran}. However, this preliminary approach only focused on the topological structure; it did not consider power system constraints, which is a necessity for power network expansion. 

In this paper, 
we build the bio-inspired robust power network design problem as a mixed-integer nonlinear programming optimization problem in \texttt{PowerModels.jl}~\cite{coffrin2018powermodels} with DC and AC power flow models respectively. Then, we compare the results to see how power flow models affect the power networks that are designed. In this paper, we focus on the design of transmission network structure and keep the generator location and load 
fixed. However, it is worth mentioning that generator placement, generation output, and load variation can also potentially affect the ecological robustness and can be considered in a similar way. A further analysis and the development of bio-inspired power networks with all these factors is our future work.

The main contributions of this paper are as follows: 
\begin{itemize}

 \item
 We formulate the bio-inpsired power networks design problem as a mixed-integer nonlinear optimization problem to directly optimize the physical network structure while satisfying power system operational constraints.
 
 \item 
 With \texttt{PowerModels.jl} \cite{coffrin2018powermodels}, we solve the bio-inspired power network design problem with DC and AC power flow models respectively and analyze the influence of different power flow models for the network design.
 
 \item
 We also compare and analyze the similarities and differences of the optimal network design in \cite{panyam2019bioT, panyam2019bioA} and in this paper with $N-x$ contingency analysis.
 
\end{itemize}

Section \ref{sec:formulation} reviews the related work of applying ecological robustness into power network design. Section \ref{model} introduces the proposed mixed-integer nonlinear optimization problem. Case studies and discussion of the bio-inspired power network design are in Section \ref{sec:case}. The conclusion and future work are in Section \ref{sec:conclusion}.

\section{Related Work} 
\label{sec:formulation}

\begin{figure}
\centering
\includegraphics[scale=0.25]{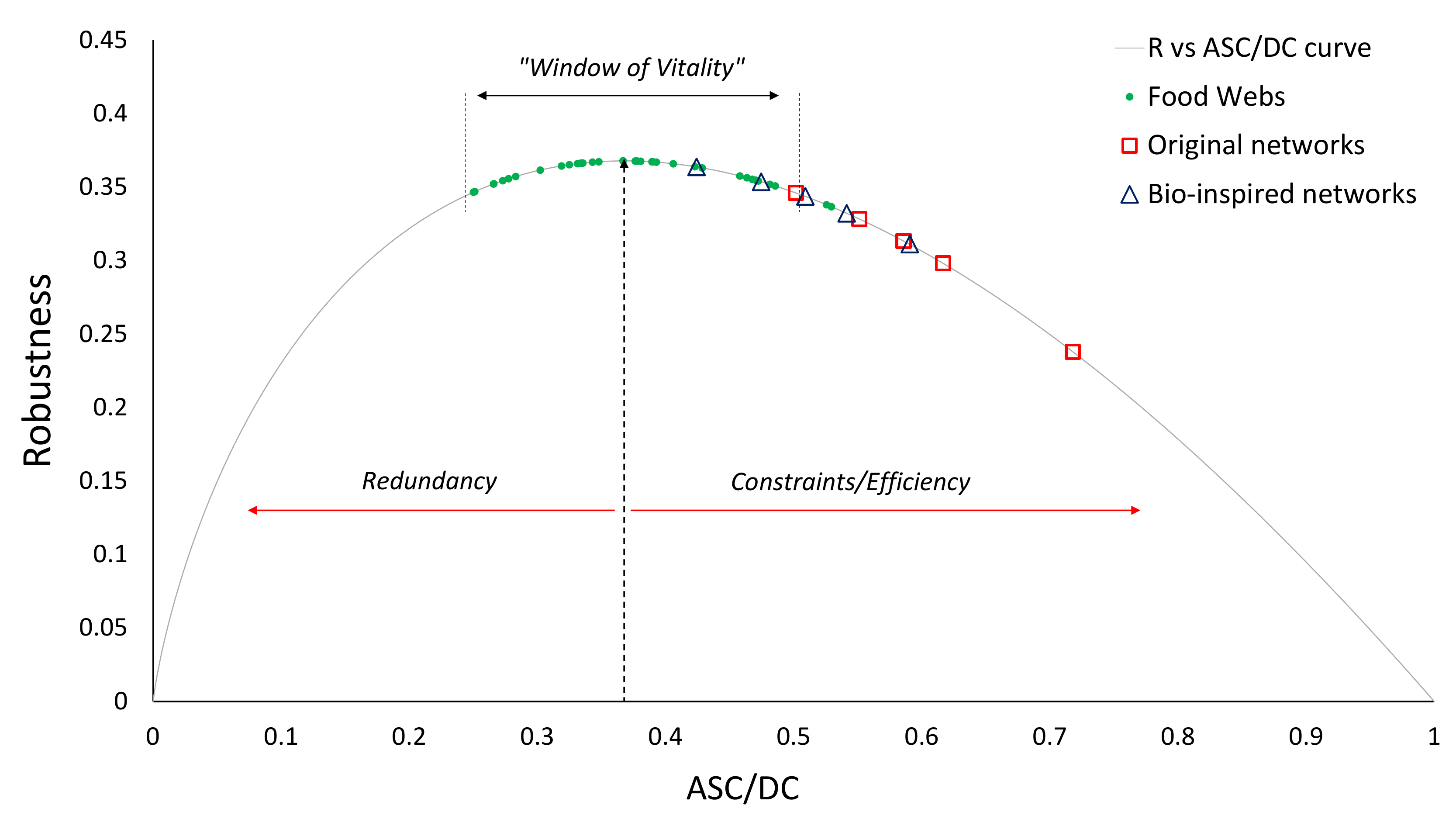}
\caption{The ecological robustness curve depicting the five grids and their bio-inspired optimized versions, as well as a set of 38 food webs.\cite{panyam2019bioA}.}
\label{fig:curve}
\vspace{-0.2cm}
\end{figure}

Our previous work has applied ecological robustness into power networks design and analysis in~\cite{panyam2019bioT, panyam2019bioA}, where we model power systems as prey-predator interactions in a food web and construct an \textit{Ecological Flow Matrix} [\textbf{T}] with real power flow for ecological robustness optimization and analysis.
The ability of bio-inspired design to improve the ecological robustness and reliability of a network is shown in \cite{panyam2019bioT, panyam2019bioA}. Fig. \ref{fig:curve} shows the ecological robustness curve for the five grids and their bio-inspired versions and a set of 38 food webs as comparison. The robustness of original power networks are much lower than the naturally robust food webs, which indicates their network structure is more efficient but less redundant. Analysis in \cite{panyam2019bioA} is performed on 5-bus, 6-bus, 7-bus, 9-bus, and 14-bus systems from \cite{powerworldcases} and the results show that the bio-optimized power networks have higher robustness with added redundancy and they are more reliable under $N-1$, $N-2$ and $N-3$ contingencies. 

\textbf{Ecological robustness}, \textit{R}, is based on the concept of \textbf{surprisal}, which is a dimensionless metric. 
It is formulated as a function of two opposing but complementary attributes: \textit{unutilized reserve capacity} and \textit{effective performance}. The optimal ecological robustness represents a unique balance between pathway \textit{efficiency} and \textit{redundancy} for a given network that can survive from more disturbances. The formulation of $R$ is shown in Eq. (\ref{eq:r}) with \textbf{total system throughput} (\textit{TSTp}) \cite{Ulanowicz1986}, \textbf{development capacity} (\textit{DC}) \cite{ulanowlcz1990symmetrical}, and \textbf{ascendency} (\textit{ASC}) \cite{Rutledge1976, Ulanowicz1980a}. 

\begin{equation}\label{eq:r}
    \textit{R} = -\bigg( \displaystyle \frac{\textit{ASC}}{\textit{DC}} \bigg) ln\bigg(\displaystyle \frac{\textit{ASC}}{\textit{DC}}\bigg)
\end{equation}

\noindent \textit{TSTp} is the system size, measured by total units of energy circulated. $DC$ is the maximum amount of aggregated uncertainty that a network can have. $ASC$ reflects the order or dependence between events. With the \textit{Ecological Flow Matrix} [\textbf{T}], the formulation of these metrics are as follows:
\begin{equation}
\label{eq:tstp}
\textit{TSTp} = \sum_{i=1}^{N+3}\sum_{j=1}^{N+3}T_{ij}
\end{equation}

\begin{equation} \label{eq:dc}
\textit{DC} = -\textit{TSTp} \sum_{i=1}^{N+3} \sum_{j=1}^{N+3} \Bigg(\displaystyle \frac{T_{ij}}{\textit{TSTp}} log \Big(\displaystyle \frac{T_{ij}}{\textit{TSTp}}\Big)\Bigg)
\end{equation}

\begin{equation} \label{eq:asc}
\textit{ASC} = -\textit{TSTp} \sum_{i=1}^{N+3} \sum_{j=1}^{N+3} \Bigg( \frac{T_{ij}}{\textit{TSTp}} log \Bigg( \displaystyle \frac{T_{ij} \textit{TSTp}} {T_{i} T_{j}} \Bigg) \Bigg)
\end{equation}

\noindent where $T_i$ and $T_j$ are the sum of all flows out of $i$ and into $j$ respectively. The \textit{Ecological Flow Matrix} [\textbf{T}] is a square ($N$+3) x ($N$+3) matrix, where \textit{N} is the number of actors, and the extra entries represent the system inputs, useful exports, and dissipation \cite{Layton2014}.

 The above formulation of \textit{R} allows the quantification of robustness as a function of pathway redundancy and efficiency, which has been shown in ecosystems to be directly related to the long-term survival of a network~\cite{Ulanowicz2009a}. In ecosystems, the maximum \textit{R} is achieved with a unique ratio of \textit{ASC/DC}, whose value is 0.367 or 1/$e$.

\section{Mixed-Integer Nonlinear Optimization Problem Formulation}\label{model}

The goal for bio-inspired power network design is to create a network that is maximally robust 
(based on the ecological definition) against large-scale hazards, while respecting real-world power system constraints.
To compute and optimize the ecological robustness for power systems, we 
focus on the real power flows and formulate the \textit{Ecological Flow Matrix} as in \cite{panyam2019bioA}. However, unlike~\cite{panyam2019bioT, panyam2019bioA}, this paper presents an optimization problem that directly optimizes the power grid physical structure instead of the \textit{Ecological Flow Matrix}. In this way, power system operational constraints, such as thermal limits, voltage magnitude limits, etc., are included.

\begin{figure*}[t]
\centering
\includegraphics[trim={1.2mm 1.2mm 1.2mm 1.2mm}, clip,width=\linewidth]{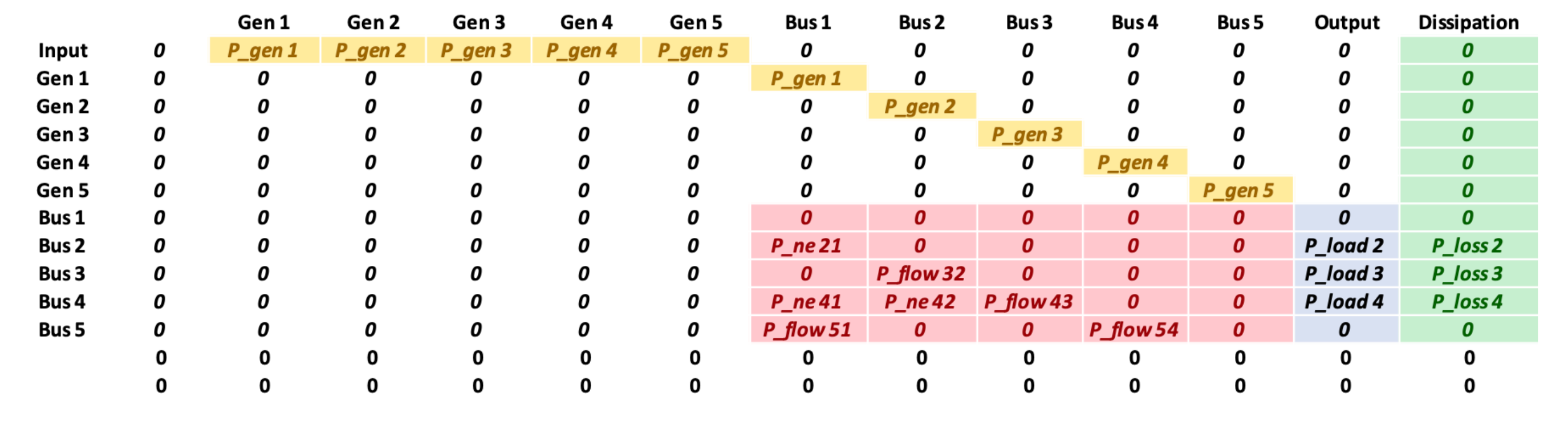}
  \caption{ An illustrative \textit{Ecological Flow Matrix} for the 5 bus network expansion case. 
  The \textit{P\_gen} $i$ is the real power output from generator $i$, which locates at the input row and the flow between generator and corresponding bus. The generators are treated as lossless, so the dissipation column for generators is 0. The \textit{P\_load} $i$ is the real power consumption at Bus $i$ and the \textit{P\_loss} $i$ is the real power loss at Bus $i$. The \textit{P\_flow} $ij$ and \textit{P\_ne} $ij$ are the real power flows at the corresponding branch and new branch.} 
  \label{5case}
  \vspace{-0.5cm}
\end{figure*}

For power grids, \textit{actors} correspond to generators and buses, \textit{system inputs} correspond to energy from generators, \textit{useful exports} are the load (consumption), and \textit{dissipation} is the losses in the system. The entries in [\textbf{T}] are notated as $T_{ij}$ and represent the directed flow from node $i$ to node $j$, corresponding to real power flows \cite{panyam2019bioA}. An illustrative \textit{Ecological Flow Matrix} using a 5-bus case from \texttt{PowerModels.jl} \cite{GitPower} transmission network expansion problem is shown in Fig. \ref{5case}. This case has 5 buses, 5 generators, and 3 loads. The loads are located at Bus 2, Bus 3, and Bus 4 respectively. There are 4 branches in the system, and 3 candidates of branches to be built. Using the \textit{Ecological Flow Matrix}, the optimal ecological robustness can guide the design of a secure and resilient network. 

To include power system operational constraints, we introduce a binary decision variable $\alpha$ for branches, whose value is either one or zero. For example, $\alpha_{ij}$ is the binary decision variable for branch from bus $i$ to bus $j$. With the proposed optimization, if the resulting $\alpha_{ij}$ equals one, it suggests to build this branch to achieve optimal ecological robustness. Otherwise, it suggests not. With the integration of power flow equations, the outputs of the generators are also decision variables. In this way, we can also optimize the operating point for the system based on ecological robustness. 

The bio-inspired robust power network design optimization with power system constraints is shown as follows:

\begin{subequations}

 \textbf{Variables}:\\ 
V\_{i} \; ($\forall$ i $\in$ N), P\_gen\_{i}\;($\forall$ i $\in$ G), \; P\_{ij} \; ($\forall$ (i,j) $\in$ B $\cup$ NB)  \\
Q\_{ij} \; ($\forall$ (i,j) $\in$ B $\cup$ NB) ,$\alpha$\_{ij} $\in$ \{0,1\} \; ($\forall$ (i,j) $\in$ NB)

\textbf{Objective}: 

\begin{equation}
 Max(R=\textit{f (\textbf{EFM}})) 
\label{7a}
\end{equation}


\textbf{Subject to}:
\begin{equation}
   v^{l}_{i} \leqslant V_{i} \leqslant v^{u}_{i} \; (\forall i \in N)   
\label{7b}
\end{equation}

\begin{equation}
 s^{l}_{ij} \leqslant \lvert S_{ij} \rvert  \leqslant s^{u}_{ij} \; (\forall (i,j) \in B \cup NB)  
\label{7c}
\end{equation}

\begin{equation}
 s\_gen^{l}_{i} \leqslant \lvert S\_gen_{i} \rvert \leqslant s\_gen^{u}_{i}  \; (\forall i \in G) 
\label{7d}
\end{equation}



\begin{equation}
\begin{split}
P\_{ij} = V_i^2[-G_{ij}] + 
V_i V_j [G_{ij} cos(\theta_{ij}) + B_{ij} sin(\theta_{ij})]\\ (\forall (i,j) \in B)
\label{7g}
\end{split}
\end{equation}

\begin{equation}
\begin{split}
P\_{ij} = \alpha_{ij}V_i^2[-G_{ij}] + 
\alpha_{ij}V_i V_j [G_{ij} cos(\theta_{ij})+ B_{ij} sin(\theta_{ij})]\\(\forall (i,j) \in NB )
\label{7h}
\end{split}
\end{equation}

\begin{equation}
\begin{split}
Q\_{ij} = V_i^2[B_{ij}] + 
V_i V_j [G_{ij} sin(\theta_{ij}) - B_{ij} cos(\theta_{ij}) ] \\ (\forall (i,j) \in B)
\label{7i}
\end{split}
\end{equation}

\begin{equation}
\begin{split}
Q\_{ij} =\alpha_{ij} V_i^2[B_{ij}] + 
\alpha_{ij}V_i V_j [G_{ij} sin(\theta_{ij}) - B_{ij} cos(\theta_{ij}) ] \\(\forall (i,j) \in NB )
\label{7j}
\end{split}
\end{equation}

\begin{equation}
P_{i} =P\_load_{i}-P\_gen_{i}= \sum_{j}P\_{ij}\; (\forall \in B \cup NB)
\label{7k}
\end{equation}

\begin{equation}
Q_{i} =Q\_load_{i}-Q\_gen_{i}= \sum_{j}Q\_{ij} \;(\forall i \in B \cup NB)
\label{7l}
\end{equation}


\begin{equation}
P\_loss_{i} =\frac{1}{2} \sum_{j} (P\_{ij}^2+Q\_{ij}^2)/(B_{ij}V_{i}^2) (\forall i \in B \cup NB)
\label{7m}
\end{equation}

\begin{equation}
\textbf{EFM}=\textbf{T}(P\_{ij},P\_{gen},P\_{load}, P\_{loss})
\label{7n}
\end{equation}

\end{subequations}\label{alg}

\noindent where \textbf{EFM} is \textit{Ecological Flow Matrix},
$B$ is the set of existing branches,
$NB$ is the set of candidates of new branches,
$N$ is the set of buses,
and $G$ is the set of generators.

The objective function (\ref{7a}) is formulated through Equation (\ref{eq:r})-(\ref{eq:asc}). The input of the objective function is a function of \textbf{EFM}, which is consisted of power flows, generation output, power loss and load of the interested power grid network as shown in Fig.\ref{5case}. 




With the binary decision variable $\alpha$ and the formulation of ecological robustness $R$, this optimization problem becomes mixed-integer nonlinear programming problem (MINLP), which is a typical class of NP-hard problem~\cite{belotti2013mixed}. To solve this problem, we build our model with \texttt{PowerModels.jl} \cite{coffrin2018powermodels, GitPower}, which provides various libraries for power system optimization problems. 

\section{Case Studies}\label{sec:case}


\texttt{PowerModels.jl} is a Julia-based~\cite{bezanson2012julia} open-source platform for comparing and evaluating different power flow formulations on actively researched optimization problems, including Optimal Power Flow (OPF), Transmission Network Expansion Planning (TNEP), Optimal Transmission Switching (OTS), etc \cite{GitPower}. In this paper, we combine the TNEP in \texttt{PowerModels.jl} with our bio-inspired power network design problem to fully utilize its existing modeling components and formulate our own objective with ecological robustness.

Based on the proposed mixed-integer nonlinear
optimization model, we build the bio-inspired robust power network design problem in \texttt{PowerModels.jl} and evaluate it for the 5-bus network expansion case from \texttt{PowerModels.jl} and a PowerWorld Simulator 6-bus case \cite{powerworldcases} using both DC and AC models.
To solve the MINLP, we use the \textit{Ipopt}~\cite{wachter2006implementation}, the nonlinear optimization solver, \textit{Juniper}~\cite{kroger2018juniper}, the nonlinear integer program solver, and \textit{Cbc}~\cite{cbc}, the mixed integer program solver. The solvers are configured to stop when the desired convergence tolerance is 
less than 10$^{-4}$ or the solver has run over 1500 seconds.

\begin{table*}[t]
\caption{Results of Mixed-Integer Optimization  for Bio-Inspired Power Network Design} 
\vspace{-.1in}
\begin{center}
\begin{tabular}{c|c|c|c|c|c|c} 
    \hline
    \hline
    \textbf{Use Case} & \textbf{Optimization Package}& \textbf{Power Flow Model} & \textbf{Solved Status} & \textbf{Optimal $R$} & \textbf{New Branch(es)}& \textbf{Time(sec)}\\ 
    \hline
5 Bus Case& \texttt{PowerModels.jl} &	DC Model&  LOCALLY SOLVED &0.349838	& 2-4, 1-4& 0.494 \\
5 Bus Case	& \texttt{PowerModels.jl} & AC Model	& INFEASIBLE	& NaN &NaN	& 0.424 \\
6 Bus Case&	 \texttt{PowerModels.jl} & DC Model	& LOCALLY SOLVED	& 0.360092 & 2-5, 3-4 &	0.716\\ 
6 Bus Case&	\texttt{PowerModels.jl} & AC Model   &INFEASIBLE	& NaN &NaN	& 1.789\\ 
6 Bus Case \cite{panyam2019bioA} & \texttt{MATLAB} &	DC Model    & LOCALLY SOLVED  & 0.353  &2-5, 1-4, 2-3, 3-4, 1-3 & 130 \\
    \hline
    \hline
\end{tabular}
\vspace{-.2in}
\end{center}
\label{tbl:result}
\end{table*}

\begin{figure}[t]
\centering
\includegraphics[trim={0mm 0mm 0mm 0mm}, clip,width=\linewidth]{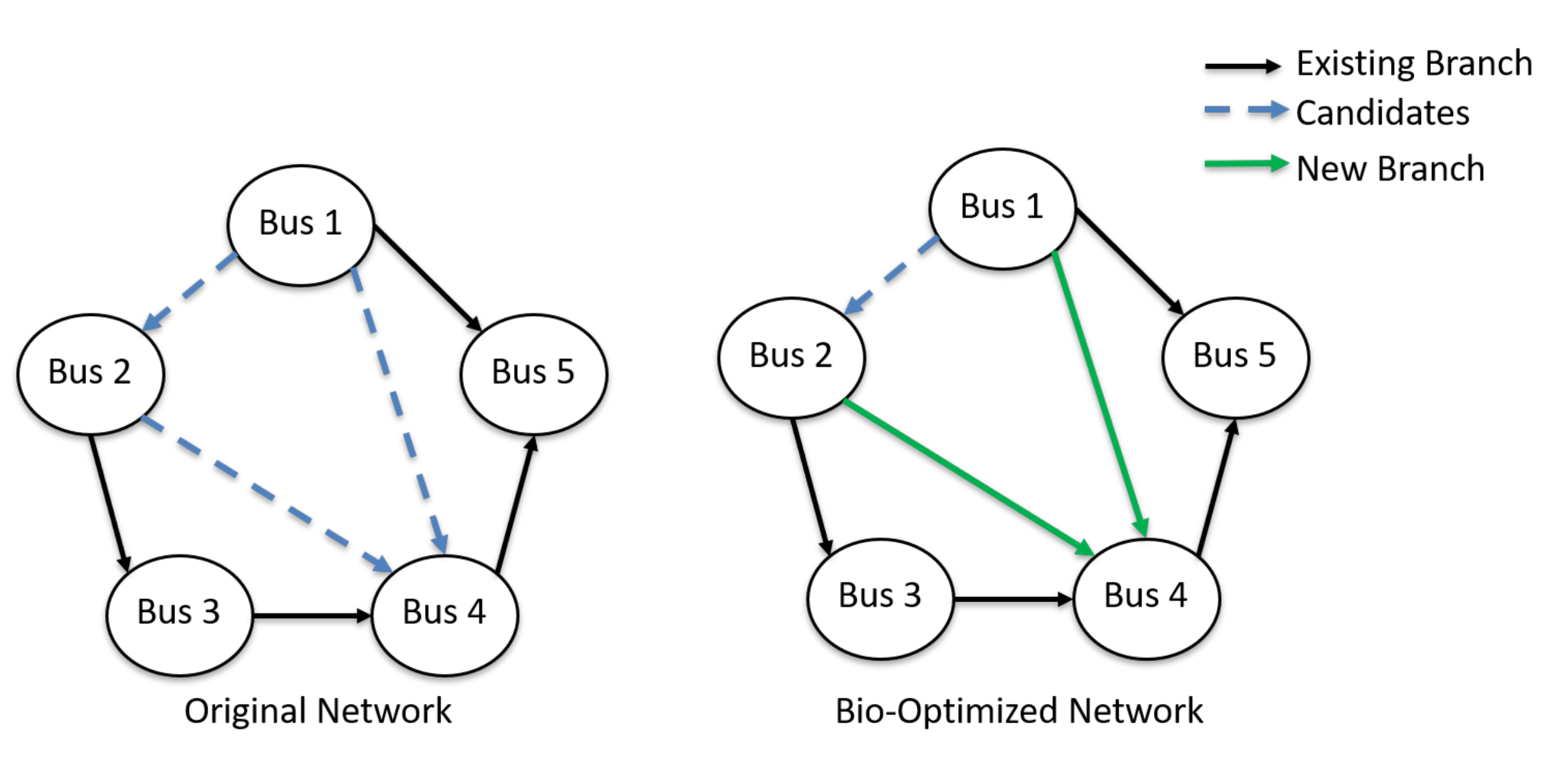}
  \caption{The network design of 5-bus case}
  \label{5casenetwork}
  \vspace{-0.2cm}
\end{figure}

\begin{figure}[t]
\centering
\includegraphics[trim={0mm 0mm 0mm 0mm}, clip,width=\linewidth]{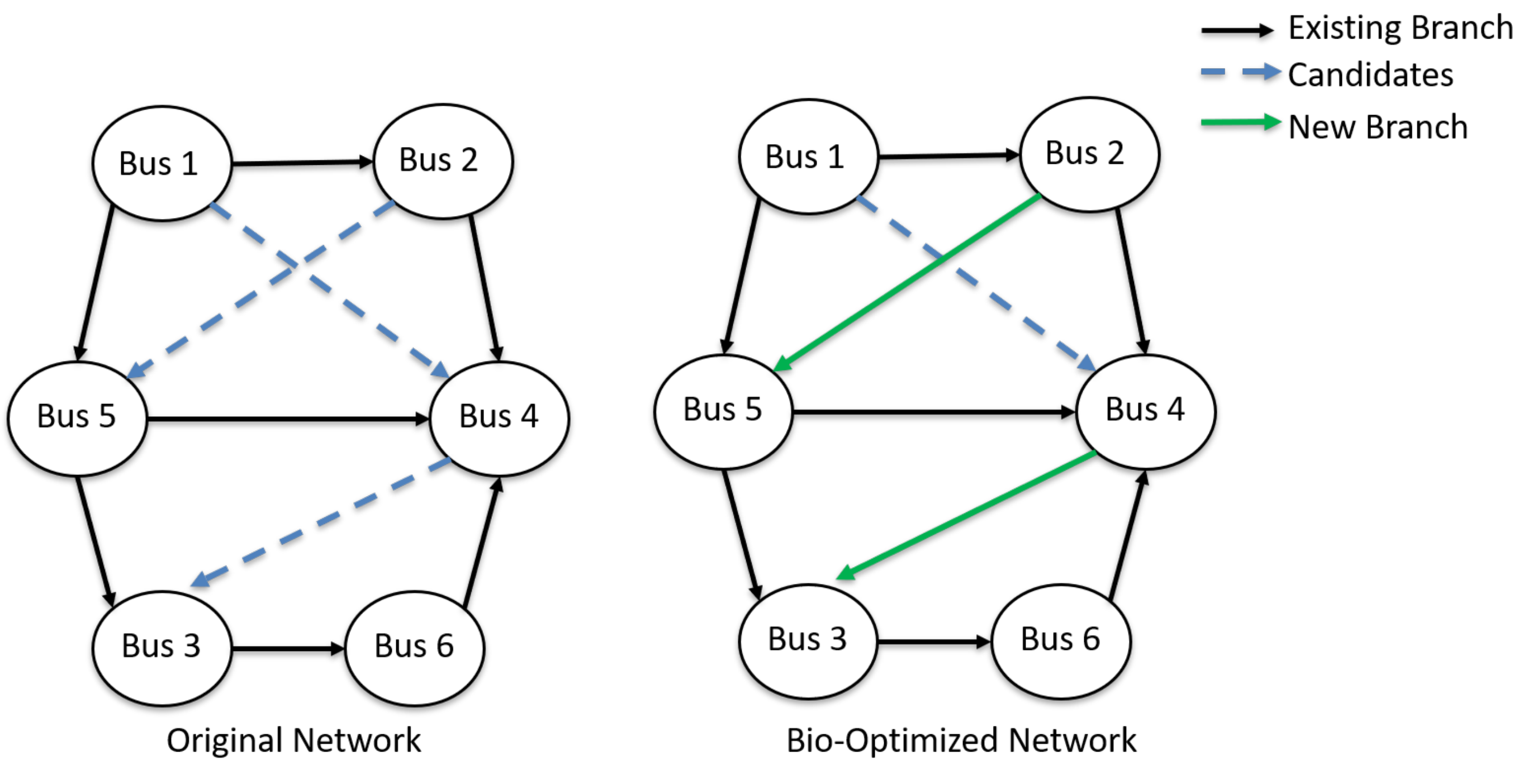}
  \caption{The network design of 6-bus case}
  \label{6casenetwork}
  \vspace{-0.2cm}
\end{figure}

In this paper, we formulate the bio-inspired power network design optimization problem with DC and AC power models respectively and apply the formulation to the 5-bus network expansion case from \texttt{PowerModels.jl}, which is under contingency with one overloaded branch and two over capacity generators due to the removal of whose \texttt{Branch 1-2} and \texttt{Branch 1-4} ($from\_bus - to\_bus$), and the 6-bus case from PowerWorld Simulator \cite{powerworldcases}, which is under normal operation. We add network expansion candidate branches to the original network to fit the cases into the proposed model. 
In our case, the branch candidates are chosen to be \texttt{Branch 1-2}, \texttt{Branch 1-4} and \texttt{Branch 2-4} in the 5-bus network expansion case as shown in Fig.~\ref{5casenetwork}. \texttt{Branch 1-2} and \texttt{Branch 1-4} are removed from the original network. \texttt{Branch 2-4} is a new branch and its parameters, such as resistance, reactance, susceptance and limit, are the same as \texttt{Branch 1-4}. The original 6-bus case is a normal power grid case with no contingency. To fit the proposed power network optimization, we select three candidates from the results of ~\cite{panyam2019bioA} in $branch.ne$ with \texttt{Branch 1-4}, \texttt{Branch 2-4} and \texttt{Branch 4-3} as shown in Fig.~\ref{6casenetwork}. The electrical parameters for the new branches are the same as the existing branches. 

The results of the MINLP bio-inspired network design are in Table~\ref{tbl:result}, and the network structures for 5-bus and 6-bus cases are shown in Fig.~\ref{5casenetwork} and Fig.~\ref{6casenetwork} respectively. It can be clearly seen that the optimal network design only chooses two more branches for both cases, and their optimal ecological robustness is much higher than the ones obtained in \cite{panyam2019bioA}. In \cite{panyam2019bioT, panyam2019bioA}, the optimization model searches all possible connections for a given network; however, the proposed optimization model pre-defines the searching space of 3 candidates, which can reduce the number of constructed branches. Besides, with the integration of power flow equations, the proposed optimization model not only guides the network design, but also adjusts the power flow dispatch to achieve the maximum robustness. Moreover, with the optimization in \cite{panyam2019bioA}, it takes 130 seconds to solve the optimal network design for the 6-bus case in \texttt{MATLAB} using $fmincon$ function with a full search. With the mixed-integer optimization in Julia, it only takes 0.716 seconds with the DC power flow model and 1.789 seconds with the AC power flow model with predefined searching space. While the two optimization problems are solved in different platforms with different solvers, constraints, and search space, making it is hard to directly compare the efficiency for them, \texttt{PowerModels.jl} demonstrates its efficiency and capability for solving emerging power system optimization problems. Moreover, the MINLP formulation provides more flexibility to choose candidates, since users can determine the candidates in advance. With geographic and economic studies, users can provide feasible and economical candidates for the model. 

The optimization is locally solved with feasible solutions for both cases with DC power flow model, but they are locally infeasible with AC power flow model. The proposed mixed-integer nonlinear optimization for bio-inspired network design is particularly complicated because the mixed-integer optimization is a NP hard problem and its objective function includes the natural logarithm, which requires the domain to be positive. With the integration of power system constraints, the feasible space for such problem is further reduced. For the 6-bus case, if we increase the number of candidates for $branch.ne$, the problem will not be solved. Thus, for different cases and power flow models, it needs further relaxation to obtain an optimal and feasible solution.

For the 6-bus case, we perform the $N-x$ contingency analysis to compare the reliability of different network designs and the results are in Table \ref{tbl:contingency}. We compare the case with the original network structure, the \texttt{PowerModel.jl} TNEP optimization structure, the bio-inspired network design from \cite{panyam2019bioA} and the proposed MINLP bio-inspired network design. The \texttt{PowerModel.jl} TNEP optimization problem constructs extra branches to ensure the system can function securely with minimum construction cost. Since the 6-bus case can function safely, the output of the 6-bus case through \texttt{PowerModel.jl} TNEP model is the same as the original one. It satisfies the TNEP model that ensures the system can function without violating any power system constraints with zero construction cost. The considered violation in this paper is the branch overflow in the system under $N-x$ contingency, which is the same as in \cite{panyam2019bioA}. From the result, it can be seen that the bio-inspired networks are more reliable than the original network and \texttt{PowerModel.jl} TNEP by encountering less number of violations under contingencies. However, compared with the bio-inspired network in \cite{panyam2019bioA}, the MINLP bio-inspired network is less reliable with a slightly higher number of violations in $N-2$ and $N-3$ contingency analysis, but the violations are still much less than the original network. From economic perspective, the MINLP bio-inspired network builds less transmission lines than the network in \cite{panyam2019bioA}, which reduces the construction cost.

\begin{table*}
\caption{$N-x$ Contingency Analysis for the 6-bus case}
\vspace{-.1in}
\begin{center}
\begin{tabular}{c|c|c|c|c} 

    \hline
    \hline
    \textbf{Network Design} &  \textbf{\# of New Branches} & \textbf{\# N-1 Violations} & \textbf{\# N-2 Violations} & \textbf{\# N-3 Violations}\\ 
    \hline
6 Bus Original Network&	0	& 9	& 122  &460	\\	
 \hline

6 Bus \texttt{PowerModel.jl} TNEP Network &	0    & 9  & 122 & 460 \\	
    \hline
6 Bus Bio-Inspired Network \cite{panyam2019bioA}  &	5    & 0  & 18 &210 \\	
 \hline
6 Bus Mixed-Integer Bio-Inspired Network &	2    & 0  & 29 & 241 \\	
    \hline
    \hline
    
\end{tabular}

\vspace{-.2in}
\end{center}
\label{tbl:contingency}
\end{table*}


\section{Conclusion and Future Work}\label{sec:conclusion}
In this paper, we present a resilience-focused power grid network design optimization with realistic power system constraints inspired by ecological networks. To integrate power system constraints, the optimization is formulated as a MINLP problem and solved in \texttt{PowerModels.jl} with both DC and AC power flow models. The DC power flow models can be solved with optimal solution, but the AC power flow models cannot find the feasible solution. A particular challenge of solving the proposed MINLP problem is how to ensure the problem can find a feasible solution under the power system constraints and natural logarithm's domain requirement. 
With the integration of power flow equations, the MINLP optimization not only strategically chooses the branches, but also adjusts the generator output to achieve the optimal ecological robustness. From the contingency analysis for the 6-bus case with different network structures, we observe the improvement of power system reliability through the mixed-integer bio-inspired network design. Even though the improvement is not as much as our previous bio-inspired network design, the branches to be built are reduced, which reduces the construction cost. Moreover, there are only three candidates to search within the proposed method, unlike the global search in our previous work, which may miss other optimal points with different candidates.

For future work, we will investigate how to select the candidates for each case and how to relax the proposed optimization problem to solve it with larger power system cases for resilient network expansion. It is also important to justify the investment of construction, based on savings that would occur during contingencies. Thus, we will focus on a detailed economic analysis of the networks being optimized to consider construction costs as well as economic lost under contingencies. The plan then is to incorporate economic parameters into the optimization formulation to result in a cost-effective robust power network design. Additionally, while generator outputs are already included decision variables, future work will also investigate the placement of generators and renewable energies for a resilient power network design.

\section*{Acknowledgment}
The authors would like to acknowledge the US Department of Energy Cybersecurity for Energy Delivery Systems program under award DE-OE0000895, the National Science Foundation under Grant 1916142 and the Texas A\&M Energy Institute for their support of this project. 

\bibliographystyle{IEEEtran}

\bibliography{hreference}

\end{document}